\documentstyle[aps,12pt,cite]{revtex}

\input epsf

\begin{document}

\title {The euclidean propagator in a model with two non-equivalent
instantons } 

\author { J. Casahorr\'an \footnote{email:javierc@posta.unizar.es}}
\address{Departamento de F\'{\i}sica Te\'orica, \\
Universidad de Zaragoza, E-50009 Zaragoza, Spain}

\maketitle

\begin{abstract}

We consider in detail how the quantum-mechanical tunneling phenomenon occurs in a
well-behaved octic potential. Our main tool will be the euclidean
propagator just evaluated between two minima of the potential at issue. For
such a purpose we resort to the standard semiclassical approximation which takes
into account the fluctuations over the instantons, i.e. the finite-action
solutions of the euclidean equation of motion. As regards the one-instanton
approach, the functional determinant associated with the so-called
stability equation is analyzed in terms of the asymptotic behaviour of
the zero-mode. The conventional ratio of determinants takes as reference
the harmonic oscillator whose frequency is the average of the two
different frequencies derived from the  minima of the potential involved
in the computation.
The second instanton of the model is studied in a similar way. The physical
effects of the multi-instanton configurations are included in this context
by means of the {\it alternate dilute-gas approximation} where the two
instantons participate to provide us with the final expression
of the propagator.  

\end{abstract}

\vfill \eject

\section {Introduction.}

The tunneling phenomenon through classically forbidden regions represents
one of the most outstanding effects in quantum theory. Starting from the
pioneering work of Polyakov on the subject \cite{po}, the semiclassical
treatment of the tunneling is usually presented via the euclidean version
of the path-integral formalism. The basis of this approach relies on the
so-called instanton calculus.
In this scheme the instantons correspond to localised
finite-action solutions of the euclidean equation of motion where the time
variable is essentially imaginary. To sum up, one finds the appropiate
classical configuration and subsequently evaluate the term associated with
the quadratic fluctuations. In doing so the functional integration itself
is solved by means of the gaussian scheme except for the zero-modes which
appear in connection with the translational invariances of the system. 
Next one introduces a set of collective coordinates so that ultimately the
gaussian integration is performed along the directions orthogonal to the 
zero-modes. In principle a functional determinant includes an infinite
product of eigenvalues so that a highly divergent result is expected. 
However one can regularize these fluctuation factors by means of the
ratio of determinants. \par

Let us describe in brief the instanton calculus for the one-dimensional
particle as can be found for instance in \cite{kl}. 
We assume that our particle moves under the action of a confining
potential $V(x)$ which yields a pure discrete spectrum of energy eigenvalues.
In addition  the minima of the
potential satisfy $V(x) = 0$.
If the particle is located at the initial time $t_i = -T/2$ at the point
$x_i$ while one finds it when $t_f = T/2$ at the point $x_f$, the
functional version of the non-relativistic quantum mechanics allows us to
write the transition amplitude in terms of a sum over all paths joining
the world points with coordinates $(-T/2, x_i)$ and $(T/2, x_f)$. 
Performing the change $t \rightarrow - i \tau$, known in the
literature as the Wick rotation, the euclidean formulation of the
path-integral reads 

\begin{equation}
<x_f\vert \exp(- H T) \vert x_i> = N(T) \int [dx]  \ 
\exp \lbrace - S_e[x(\tau)] \rbrace
\label{eq:1}
\end{equation}

\noindent where $H$ represents the hamiltonian of the model, the
factor $N(T)$ serves to normalize the amplitude conveniently while
$[dx]$ indicates the integration over all functions which fulfil the
boundary conditions at issue. As usual the euclidean
action $S_e$ corresponds to 

\begin{equation}
S_e =  \int_{-T/2}^{T/2} \left[  {{1} \over {2}} 
\left( {{dx} \over {d\tau}} \right)^2 + V(x) \right]  \ d\tau
\label{eq:2}
\end{equation}

\noindent whenever the mass of the particle is set equal to unity.
In the following we take care of the octic potential $V(x)$ given by

\begin{equation}
V(x) = {{\omega^2 } \over {2}}  (x^2 - 1)^2 (x^2 - 4)^2
\label{eq:3}
\end{equation}

\noindent whose appearance can be seen in fig.1.
Going to the regime in which  $\omega^2 \gg 1$ the energy
ba\-rri\-ers are high enough to decompose the physical system into a sum of 
independent harmonic oscillators. In doing so the particle can execute
small oscillations around each minima of the potential located at
$x = \pm 1$ and $\tilde{x} = \pm 2$. As usual the second derivative of the
potential just evaluated at these points, i.e. $V''(x = \pm 1) = 36 \  \omega^2$
and $V''(\tilde{x} = \pm 2) = 144 \  \omega^2$ characterizes the respective frequencies
of the harmonic oscillators at issue. \par

\begin{figure}[htb]
\centerline{\epsfbox[0 0 300 200]{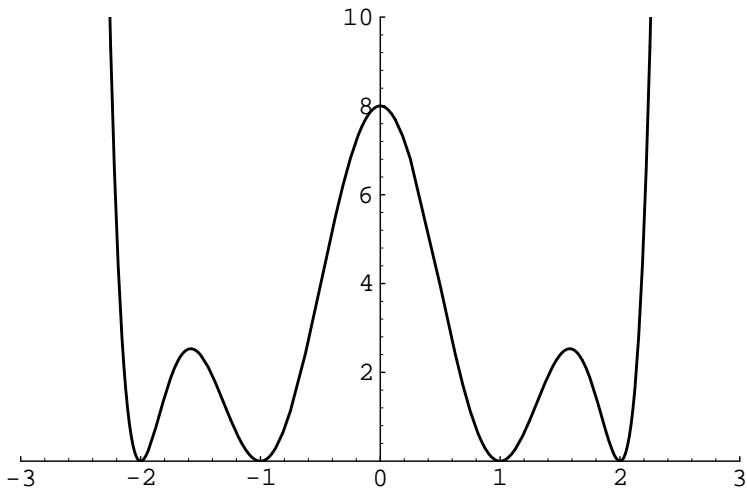}}
\caption{Profile of the potential.}
\label{Figure 1}
\end{figure}

In terms of the discrete symmetry $x \rightarrow - x$ which the potential
$V(x)$ enjoys, one notices how the four minima are non-equivalent as a whole
since no connection is possible between the two sets represented by
$x = \pm 1$ and $\tilde {x} = \pm 2$. Next we would like to make the description
of the tunneling phenomenon to explain how the symmetry cannot appear
spontaneously broken at quantum level. In such a case the expectation value
of the coordinate $x$  evaluated for the ground-state is zero as corresponds
to the even character of the potential $V(x)$. \par 

\section{The one-instanton contribution.}

In this section we carry things further to describe the tunneling phenomenon
according to the tools of the euclidean version of the path-integral. At first
glance it seems physically relevant to take into account the transition between
the points $x_{i} = 1$ and $x_{f} = 2$.
In doing so we need the explicit form of the topological configuration
with  $x_{i} = 1$ at $t_{i} = -T/2$ while
$x_{f} =  2$ when $t_{f} = T/2$. 
It is customarily assumed that $T \rightarrow \infty$ since
the explicit solution of the problem is much more complicated for finite $T$.
Fortunatelly the difference is so small that can be ignored mainly because
we are interested in such a  limit to obtain information about the energy
of the first levels of the model. To get the instanton $x_{c1}(\tau)$ which
connects the points $x_{i} = 1$ and $x_{f} = 2$ with infinite euclidean
time, we can resort to the well-grounded Bogomol'nyi condition \cite{bo}.
The problem is solved just by integration of a first-order differential equation
according to the zero-energy condition for the motion of a single particle
under the action of $- V(x)$. To sum up 

\begin{equation}
x_{c1}(\tau) =  2 \cos \left[ {{\pi} \over {3}} - {{1} \over {3}} \arccos
\left({{e^{- 12 \omega (\tau - \tau_{c})} - 1}
\over {e^{- 12 \omega (\tau - \tau_{c})} + 1}} \right) \right] 
\label{eq:4}
\end{equation}

\noindent where the parameter $\tau_c$ indicates the point at which the instanton
makes the jump.   As usual equivalent solutions 
are  obtained by means of the transformations $\tau \rightarrow - \tau$ and
$x_{c1}(\tau) \rightarrow - x_{c1}(\tau)$.
It may be interesting at this point to remind how the instanton procedure
allows only the connection between adjoint minima of the potential. In this
regard we notice the existence of a second instanton just interpolating
between $x_{i} = -1$ and $x_{f} = 1$ (more on this later). 
Going back to $x_{c1}(\tau)$, the classical euclidean action $S_{1}$ associated
with a such configuration is computed according to (\ref{eq:2}) so that
$S_{1} = 22 \omega/15$. Now the conventional description of the one-instanton
amplitude between $x_{i} = 1$ and $x_{f} = 2$ takes over

\begin{eqnarray*}
<x_f = 2\vert \exp(- H T) \vert x_i = 1> = N(T) \left\{ Det \left[
- {{d^2} \over {d\tau^2}}  +
 \nu^2  \right]\right\}^{-1/2}   
\end{eqnarray*}
\begin{equation}
\left\{{{Det \left[- (d^2/d\tau^2) + V^{\prime \prime}[x_{c1}(\tau)] \right]} \over
{Det \left[- (d^2/d\tau^2) + \nu^2 \right]}}\right\}^{- 1/2} \ 
\exp(-S_{1})
\label{eq:5}
\end{equation}

\noindent where we have multiplied and divided by the determinant of a generic
harmonic oscillator of frequency $\nu$. This regularization term can be
interpreted as a new amplitude given by

\begin{equation}
<x_f = 0\vert \exp(- H_{ho} T) \vert x_i = 0> = N(T) \left\{ Det \left[
-  {{d^2} \over {d\tau^2}}  +
 \nu^2  \right]\right\}^{-1/2}
\label{eq:6}
\end{equation}

Fortunatelly the explicit evaluation of (\ref{eq:6}) is made according to
the procedure exposed in \cite{ra}. In short

\begin{equation}
<x_f = 0\vert \exp(- H_{ho} T) \vert x_i = 0> = \left({{\nu}
\over {\pi}}\right)^{1/2} \
\left(2 \sinh \nu T \right)^{-1/2}
\label{eq:7}
\end{equation}

Going back to the determinant of the stability equation over the instanton
$x_{c1}(\tau)$, the existence of a zero-mode $x_{o}(\tau)$ requires the
introduction of a collective coordinate. From a physical point of view
this zero eigenvalue comes by no surprise since it reflects the translational
invariance of the system as a whole. In other words, there is one direction
in the functional space of the second variations which results incapable
of changing the action. Including the right factor of normalization the
explicit form of the zero-mode $x_{o}(\tau)$ corresponds to the derivative 
of the topological configuration, i.e.

\begin{equation}
x_{o}(\tau) = {{1} \over {\sqrt{S_{1}}}} {{dx_{c1}} \over {d\tau}}
\label{eq:8}
\end{equation}

In addition the integral over the zero-mode itself becomes equivalent to the
integration over the center of the instanton $\tau_{c}$. When this change of
variables is incorporated our ratio of determinants  corresponds to \cite{kl}

\begin{eqnarray*}
\left\{{{Det \left[- (d^2/d\tau^2) + V^{\prime \prime}[x_{c1}(\tau)] \right]} \over
{Det \left[- (d^2/d\tau^2) + \nu^2 \right]}}\right\}^{- 1/2} = 
 \end{eqnarray*}
\begin{equation}
\left\{{{Det^{\prime} \left[- (d^2/d\tau^2) + V^{\prime \prime}[x_{c1}(\tau)] \right]} \over
{Det \left[- (d^2/d\tau^2) + \nu^2 \right]}}\right\}^{- 1/2} \
\sqrt{{{S_{1}} \over {2 \pi}}} \ d\tau_{c}
\label{eq:9}
\end{equation}

\noindent where $Det^{\prime}$ stands for the so-called reduced determinant
once the zero-mode has been removed.
To make an explicit computation
of the quotient of determinants we take advantage of the Gelfand-Yaglom
where only the knowledge of the large-$\tau$ behaviour of the
classical solution $x_{c1}(\tau)$ is necessary \cite{gy}. Starting
from $\hat{O}$ and $\hat{P}$, which represent a couple
of se\-cond order differential operators whose eigenfunctions
vanish at the boundary, the quotient of determinants is given in terms of the
respective zero-energy solutions $f_{o}(\tau)$ and $g_{o}(\tau)$ according to

\begin{equation}
{{Det \hat{O}} \over {Det \hat{P}}} = {{f_{o}(T/2)} \over {g_{o}(T/2)}}
\label{eq:10}
\end{equation}

\noindent whenever the eigenfunctions at issue fulfil the initial conditions

\begin{equation}
f_{o}(-T/2) = g_{o}(-T/2) = 0, \ \ \ {{df_{o}} \over {d\tau}} (-T/2) =
{{dg_{o}} \over {d\tau}} (-T/2) = 1
\label{eq:11}
\end{equation}

The zero-mode $g_{o}(\tau)$ associated with the harmonic oscillator
of frequency $\nu$ corresponds to

\begin{equation}
g_{o}(\tau) = {{1} \over {\nu}} \  \sinh [\nu (\tau + T/2)]
\label{eq:12}
\end{equation}

\noindent so that now we only need the  form of the solution $f_{o}(\tau)$ 
associated with the topological configuration written in (\ref{eq:4}).
From the aforementioned $x_{o}(\tau)$ zero-mode we can write a second solution
$y_{o}(\tau)$ given by 

\begin{equation}
y_{o}(\tau) = x_{o}(\tau) \ \int_{0}^{\tau} {{ds} \over {x_{o}^{2}(s)}}
\label{eq:13}
\end{equation}

Accordingly we may summarize the asymptotic behaviour of $x_{o}(\tau)$ and
$y_{o}(\tau)$ as follows

\begin{equation}
x_{o}(\tau) \sim \left\{ \matrix{C \exp (- 12 \omega \tau) \ \ \ if \ \ 
\tau \rightarrow \infty \cr D \exp ( 6 \omega \tau) \ \ \ \  if \ \ 
\tau \rightarrow - \infty \cr} \right.
\label{eq:14}
\end{equation}

\begin{equation}
y_{o}(\tau) \sim \left\{ \matrix{ \exp ( 12 \omega \tau)/24 \omega C \ \ \ if \ \
\tau \rightarrow \infty \cr - \exp (- 6 \omega \tau)/12 \omega D \ \ \ \  if \ \
\tau \rightarrow - \infty \cr} \right.
\label{eq:15}
\end{equation}

\noindent where the constants $C$ and $D$ derive from the explicit form of 
the derivative of (\ref{eq:4}). Now we investigate the particular
solution $f_{o}(\tau)$ which is the one we are really interested in. Starting
from the linear combination of $x_{o}(\tau)$ and $y_{o}(\tau)$ given by

\begin{equation}
f_{o}(\tau) = A x_{o}(\tau) + B y_{o}(\tau)
\label{eq:16}
\end{equation}

\noindent the incorporation of the initial conditions at issue leads us to

\begin{equation}
f_{o}(\tau) = x_{o}(- T/2) y_{o}(\tau) - y_{o}(- T/2) x_{o}(\tau)
\label{eq:17}
\end{equation}

From this expression, which is exact, we can extract the asymptotic behaviour
of $f_{o}(\tau)$, i.e.

\begin{equation}
f_{o}(T/2) \sim {{D} \over {24 \omega C}} \ exp(3 \omega T) \ \ \ \  if  \ \ 
T \rightarrow \infty
\label{eq:18}
\end{equation}

Now we need to consider in detail the lowest eigenvalue of the stability
equation to obtain the right value of the quotient of determinants. From
a physical point of view we can explain the situation as follows: the
derivative of the topological solution does not quite satisfy the boundary
conditions for the interval $(- T/2,T/2)$. When enforcing such a behaviour,
the eigenstate is compressed and the energy shifted slightly upwards. In doing
so the zero-mode $x_{o}(\tau)$ is substituted for the $f_{\lambda}(\tau)$ which
corresponds to 

\begin{equation}
- {{d^2 f_{\lambda}(\tau)} \over {d\tau^2}}  +
 V^{\prime \prime}[x_{c1}(\tau)] f_{\lambda}(\tau) = \lambda f_{\lambda}(\tau)
\label{eq:19}
\end{equation}

\noindent whenever

\begin{equation}
f_{\lambda}(- T/2) = f_{\lambda}(T/2) = 0
\label{eq:20}
\end{equation}

To lowest order in perturbation theory we obtain that

\begin{equation}
f_{\lambda}(\tau) \sim  f_{o}(\tau) + \left.\lambda \ {{df_{\lambda}} 
\over {d\lambda}} \right |_{\lambda = 0}
\label{eq:21}
\end{equation}

\noindent so that ultimately we can write that

\begin{equation}
f_{\lambda}(\tau) =  f_{o}(\tau) + \lambda \int_{-T/2}^{\tau}
[x_{o}(\tau) y_{o}(s) - y_{o}(\tau) x_{o}(s)] \ f_{o}(s) \ ds
\label{eq:22}
\end{equation}

The asymptotic behaviour of $f^{o}(\tau)$, $x_{o}(\tau)$ and $y_{o}(\tau)$,
together with the condition $f_{\lambda}(T/2) = 0$ provides us with the
lowest eigenvalue $\lambda$, namely

\begin{equation}
\lambda = 12 \omega D^2 \exp(- 6 \omega T)
\label{eq:23}
\end{equation}

The final expression of the quotient of determinants requires a choice
for the parameter $\nu$ so that ultimately the frequency of 
the harmonic oscillator of reference is the average of the frequencies over the
non-equivalent minima located at $x = 1$ and  $\tilde{x} = 2$
In other words $\nu = 9 \omega$. It may be interesting at this point to remark the
differences with the well-grounded double-well model where the two minima of
the potential are equivalent so that the aforementioned average is not
necessary. However in this case the Gelfand-Yaglom method fixes  the
frequency $\nu$ and subsequently the ratio of determinants is finite.  
Going back to the explicit form of $x_{c1}(\tau)$ (see (\ref{eq:4}))
we find 

\begin{equation}
C = {{4 \sqrt{3} \omega} \over {\sqrt{S_{1}}}}, \ \ \ \  
D = {{16 \omega} \over {3 \sqrt{S_{1}}}}
\label{eq:24}
\end{equation}
 
Armed with this information we can write
the one-instanton amplitude between the points $x_{i} = 1$ and $x_{f} = 2$ 
according to 

\begin{eqnarray*}
<x_f = 2\vert \exp(- H T) \vert x_i = 1> = 
\end{eqnarray*}
\begin{equation}
 \left({{9 \omega} \over { \pi}}\right)^{1/2} \
\left(2 \sinh 9 \omega T \right)^{-1/2} \  \sqrt{S_{1}} \  
K_{1} \ 
\exp(-S_{1}) \  \omega \  d\tau_{c} 
\label{eq:25}
\end{equation}

\noindent where $K_{1}$ stands for a numerical factor given by

\begin{equation}
K_{1} = 16 \sqrt{{{15 \sqrt{3}} \over {11 \pi}}}
\label{eq:26}
\end{equation}

Apart from the first factor, which represents the contribution of the
harmonic oscillator of reference, we get a transition amplitude just
depending on the point $\tau_{c}$ at which the instanton makes  the jump.
According to the values of the interval $T$ the result seems plausible
whenever

\begin{equation}
\sqrt{S_{1}} \  
K_{1} \ 
\exp(-S_{1}) \  \omega \  T \ll 1
\label{eq:27}
\end{equation}

\noindent a nonsense condition if $T$ is large enough. However in this
regime we can accommodate configurations constructed of instantons and
anti-instantons which mimic the behaviour of a trajectory strictly derived
from the euclidean equation of motion. In doing so we get an additional bonus
since the integration over the centers of the string of instantons and
anti-instantons is performed in a systematic way. \par

To finish this section we sketch the situation as far as the second instanton
of the model concerns. For such a purpose we take into account the
one-instanton amplitude between $x_{i} = - 1$ and $x_{f} =  1$ which is based on
the topological configuration $x_{c2}(\tau)$ 

\begin{equation}
x_{c2}(\tau) =  2 \cos \left[ {{\pi} \over {3}} + {{1} \over {3}} \arccos
\left({{e^{ 12 \omega \tau} - 1}
\over {e^{ 12 \omega \tau} + 1}} \right) \right] 
\label{eq:28}
\end{equation}

\noindent whose classical euclidean action  corresponds to
$S_{2} = 76 \omega/5$.
As expected the second instanton reminds the case of the
double-well potential since connects equivalent minima of the potential. In doing
so the $x_{o}(\tau)$, $y_{o}(\tau)$ are  symmetric so that the application
of the Gelfand-Yaglom method is straightforward.
The explicit form of the ratio of determinants at issue should be

\begin{equation}
\left\{{{Det' \left[- (d^2/d\tau^2) + V^{\prime \prime}[x_{c2}(\tau)] \right]} \over
{Det \left[- (d^2/d\tau^2) + 36 \  \omega^2 \right]}}\right\}^{- 1/2} = 
\sqrt{S_{2}} \  
K_{2} \ 
\omega \  d\tau_{c} 
\label{eq:29}
\end{equation}

\noindent where $K_{2}$ is given by

\begin{equation}
K_{2} = 12 \sqrt{{{15 } \over {38 \pi}}}
\label{eq:30}
\end{equation}

\section {The alternate dilute-gas approximation.}

As all the above calculations were carried out over a single instanton, it remains
to discuss the complete amplitude which incorporates the physical effect of a
string of instantons and anti-instantons along the $\tau$ axis. The octic potential
represents in this context a more complicated case since we need to include the
whole scheme of non-equivalent instantons.
It is customarily assumed that these combinations of topological solutions
represent no strong deviations of the trajectories  derived from the
euclidean equation of motion without any kind of approximation. We wish to compute
the functional integral by summing over all such configurations, with
$n$ instantons and anti-instantons centered at points
$\tau_1,...,\tau_n$ whenever

\begin{equation}
-{{T} \over {2}} < \tau_1 < ... < \tau_n < {{T} \over {2}}
\label{eq:31}
\end{equation}

Being narrow enough the regions where the instantons (anti-instantons) make the
jump, the action of the
proposed path is almost extremal. We can carry things further and
assume that the action of the string of instantons and anti-instantons is
given by the sum of the $n$ individual actions. This scheme is well-known 
in the literature where it appears with the name of dilute-gas approximation
\cite{co}. In addition the translational degrees of freedom yield an integral 
of the form

\begin{equation}
\int_{-T/2}^{T/2} \omega d\tau_n \
\int_{-T/2}^{\tau_n} \omega d\tau_{n - 1} ... 
\int_{-T/2}^{\tau_2} \omega d\tau_1 = {{(\omega T)^n} \over {n!}} 
\label{eq:32}
\end{equation}

When evaluating the transition amplitude between $x_{i} = 1$ and $x_{f} = 2$
the total number $n$ of topological configurations must be odd. As a matter
of fact we can split $n$ (odd) into the sum of two contributions $n_{1}$ (odd)
and $n_{2}$ (even) which represent the different possibilities associated with
the existence of non-equivalent instantons. Accordingly we have $n_{1}$
topological configurations just interpolating between $x = 1$ and $\tilde{x} = 2$
or $x = - 1$ and $\tilde{x} = - 2$. 
Of course identical situation appears in connection with $n_{2}$
where now the initial and final points of the trip are  $x = \pm 1$. 
In addition we need to include a combinatorial factor $F$ to count the different
possibilities that we have of distributing the $n$ instantons.
Except for the last step which
corresponds to the instanton analyzed in the previous section, we deal with
a closed path of topological configurations starting and coming back to the
point $x = 1$. As regards the instantons (anti-instantons) belonging to the
first type we observe the formation of pairs in a systematic way due to the
location of the four minima of the potential along the real axis.
In doing so we have $(n_{1} - 1)/2 + n_{2}$ holes to fill bearing in mind
that once the $(n_{1} - 1)/2$ pairs of instantons and anti-instantons are
distributed no freedom at all remains to locate the topological configurations
associated with $n_{2}$. To sum up 

\begin{equation}
F = {(n_{1} - 1)/2 + n_{2} \choose (n_{1} - 1)/2}
\label{eq:33}
\end{equation}

As regards the quadratic fluctuations over the $n$ topological configurations
at issue the alternate dilute-gas approximation means that

\begin{eqnarray*}
\left\{{{Det^{\prime} \left[- (d^2/d\tau^2) + V^{\prime \prime}[x_{c1}(\tau)] \right]} \over
{Det \left[- (d^2/d\tau^2) + 81 \omega^2 \right]}}\right\}^{- 1/2} \longrightarrow
\end{eqnarray*}
\begin{equation}
\left[\left\{{{Det^{\prime} \left[- (d^2/d\tau^2) + V^{\prime \prime}[x_{c1}(\tau)]
\right]} \over
{Det \left[- (d^2/d\tau^2) + 81 \omega^2 \right]}}\right\}^{- 1/2}\right]^{n_{1}} \ 
\left[\left\{{{Det^{\prime} \left[- (d^2/d\tau^2) + V^{\prime \prime}[x_{c2}(\tau)]
\right]} \over
{Det \left[- (d^2/d\tau^2) + 36 \omega^2 \right]}}\right\}^{- 1/2}\right]^{n_{2}}  
\label{eq:34}
\end{equation}

With all this information we can discuss the complete transition amplitude we
are looking for once for simplicity we introduce the so-called instanton
density, i.e.

\begin{equation}
d_{i} = \sqrt{S_{i}} \ K_{i} \ exp(-S_{i}), \ \ \ i = 1, 2
\label{eq:35}
\end{equation}

In other words

\begin{eqnarray*}
<x_f = 2\vert \exp(- H T) \vert x_i = 1> = 
\left({{9 \omega} \over { \pi}}\right)^{1/2} \
\left(2 \sinh 9 \omega T \right)^{-1/2}
\end{eqnarray*}
\begin{equation}
\sum_{n_{1},n_{2}}
\left[ d_{1} \   \omega \  T \right]^{n_{1}} \
\left[ d_{2} \   \omega \  T \right]^{n_{2}} \
{{F} \over {n!}}
\label{eq:36}
\end{equation}

The sum $S$ of (\ref{eq:36}) can be written in terms of 

\begin{equation}
n = 2 r + 1, \ \ \ \ \  r = 0, 1, ... 
\label{eq:37}
\end{equation}

\begin{equation}
n_{2} = 2 q, \ \ \ \ \  q = 0, 1, ...
\label{eq:38}
\end{equation}

\noindent so that

\begin{equation}
S = \sum_{r = 0}^{\infty} \ \sum_{q = 0}^{r} \ 
{{(r + q)!} \over {(r - q)! \ (2q)!}} \
{{\left[ d_{1} \   \omega \  T \right]^{2(r-q)+1} \ 
\left[ d_{2} \   \omega \  T \right]^{2q}} \over {(2r + 1)!}}
\label{eq:39}
\end{equation}

On the other hand the best way of organizing this double sum should be
the following

\begin{equation}
S = \sum_{r = 0}^{\infty} \ {{\left[ d_{1} \   \omega \  T \right]^{2r+1}}
\over {(2r+1)!}} \
\sum_{q = 0}^{r} \ 
{r+q \choose r-q} \ (d_{2}/d_{1})^{2q}
\label{eq:40}
\end{equation}

Next we can handle the sum $\tilde{S}$ concerning the variable $q$ taking advantage of 
\cite{pb}

\begin{equation}
\sum_{q = 0}^{r} \ (-1)^{q} {r+q \choose 2q} x^{2q} = 
\sec [\arcsin (x/2)] \ \cos [(2 r + 1) \arcsin (x/2)]
\label{eq:41}
\end{equation}

\noindent including the transformation $x \rightarrow i x$ to obtain that

\begin{equation}
\tilde{S} = {{\cosh [(2 r + 1) \  \arg\sinh (s/2)]} \over
{\cosh [\arg\sinh (s/2)]}}
\label{eq:42}
\end{equation}

\noindent where $s$ stands for the relative instanton density given by
$s =d_{2}/ d_{1}$. In terms of a new variable $z$ defined as

\begin{equation}
z = \arg\sinh (s/2)
\label{eq:43}
\end{equation}

\noindent it is the case that a typical value of $r$ provides us with
the final expression for $\tilde{S}$, i.e.

\begin{equation}
\tilde{S} = {{exp[(2 r + 1) z]} \over
{\sqrt{4 + s^2}}}
\label{eq:44}
\end{equation}

Now it suffices to introduce this result in (\ref{eq:40}) so that

\begin{equation}
{S} = {{\sinh [d_{1} \omega T \exp(z)]} \over
{\sqrt{4 + s^2}}}
\label{eq:45}
\end{equation}

Next we can collect these partial results to write the final expression
for the complete euclidean transition amplitude between the points
$x_{i} = 1$ and $x_{f} = 2$, namely

\begin{eqnarray*}
<x_f = 2\vert \exp(- H T) \vert x_i = 1> = 
\end{eqnarray*}
\begin{equation}
 \left({{9 \omega} \over { \pi}}\right)^{1/2} \
\left(2 \sinh 9 \omega T \right)^{-1/2} \
{{\sinh [d_{1} \  \omega \  T \ \exp(z)]} \over
{\sqrt{4 + s^2}}}
\label{eq:46}
\end{equation}

Our approach provides a new way of dealing with quantum-mechanical models
which exhibit a more complicated structure of non-equivalent classical vacua
in comparison with the well-grounded cases of the double-well or periodic
sine-Gordon potentials where the equivalence of all the minima of $V(x)$ is
taken for granted \cite{co}. As regards the octic potential the topological
solutions of the system inherit the property of non-equivalence.
In any case the quantum fluctuations can be evaluated by means of the
Gelfand-Yaglom method. The global effect of the multi-instanton configurations is
discussed via the different kinds of instantons that take part to provide us
with the final expression of the euclidean propagator.

\vfill \eject

\end{document}